\documentclass[showpacs,preprintnumbers,amsmath,reprint,aps,pra,twocolumn]{revtex4-1}

\usepackage{bm}
\usepackage{graphicx} 
\usepackage{color}
\usepackage{amsmath}
\usepackage{slashed}

\begin{document}
 
\title{Effective Single Photon Decay Mode of Positronium Decay via Electroweak Interactions }

\author{Jes\'{u}s P\'{e}rez-R\'{i}os}

\affiliation{Department of Physics and Astronomy, 
Purdue University,  47907 West Lafayette, IN, USA}

\author{Sherwin T. Love }

\affiliation{Department of Physics and Astronomy, 
Purdue University,  47907 West Lafayette, IN, USA}

\date{\today}

\begin{abstract}
We consider the decay of positronium to a neutrino-antineutrino accompanied by a 
single photon. Since the neutrino pair go undetected, this appears as a 
single photon decay of positronium. These decay channel are mediated through
 the exchange of the massive $W$ and $Z$ vector bosons  of  
the electroweak interaction. After summing over the various neutrino 
channels, the standard model calculation yields the rate for such a single photon
 decay process of $\Gamma_{Ps \rightarrow \gamma}$ = 1.72 $\times 10^{-19}$ s$^{-1}$.

\end{abstract}

\maketitle

\section{Introduction}

Positronium (Ps) is a metastable leptonic atom~\cite{Positron-Physics} composed of an 
electron and its antiparticle, the positron. This exotic atom resembles in many ways 
the  hydrogen atom, but with bound state energy levels appearing  
at half of the energies due to its reduced mass being half of the electron mass. The existence of 
Ps was firstly postulated by Mohorovicic in 1934~\cite{Rich-1981} and finally 
observed in 1951 by Deutsch~\cite{Deutsch-1951}. As in the 
case of the hydrogen atom, two Ps can form a molecular 
bound state, leading to a hydrogen-like molecule but with 
a leptonic nature Ps$_{2}$, which was predicted in the 40's 
by Wheeler~\cite{Wheeler-1946}, and recently 
observed~\cite{Cassidy-2007}. Different decay channels 
of both positronium and molecular positronium Ps$_{2}$ have been addressed in the 
framework of QED~\cite{Frolov-2009,Puchalski-2008,JPR-2015,Ps_decay}.

Just as the case with the Hydrogen atom, Ps possesses two 
different spin states, a singlet state which is called parapositronium 
(p-Ps), and the triplet state called orthopositronium (o-Ps). Within 
the strict confines  
quantum electrodynamics (QED), which preserves charge conjugation symmetry, the p-Ps state can only 
decay  into an even number of photons, while 
o-Ps decays to an odd number of photons~\cite{Peskin}. 
The vast majority of the decay modes of Ps which are allowed within pure  
QED, both the dominant modes as well as more exotic decay channels, have 
been computed~\cite{Ps_decay}. For the dominant modes the calculations have been extended to include higher order QED radiative corrections. 
The inclusion of the  
electroweak interaction which are mediated by the exchange of the
 massive vector $W$ and $Z$ bosons allows for additional  
decay modes such as those involving neutrinos which are absent in 
pure QED analysis. For instance, o-Ps can decay into two neutrinos 
leading to an invisible channel~\cite{Govaerts-1996,Ps_decay} which 
has been experimentally investigated~\cite{Badertscher-2007}. However, 
the influence of the electroweak interaction into the decay modes 
of p-Ps remain unexplored.

In this work, we report on the calculation of the  decay 
mode of p-Ps into a neutrino-antineutrino plus a single photon. 
Since the neutrino pair goes undetected, 
this appears experimentally as a single photon decay channel. The 
studied decay mode involves the exchange of 
massive vector bosons of the electro-weak interaction and as such
cannot arise within pure QED. The energy of the photon emitted in 
this mode exhibits  a continuous spectra of energy between zero and 
m$_{e}$. The process can shed some additional light on the influence 
of the electroweak interaction in leptonic bound states, as well 
constituting another novel decay channel for the completeness of the 
decay chart of Ps.  
  
\section{Single photon decay mode of Ps}

The annihilation of Ps into a neutrino-antineutrino plus a single photon is a consequence of the 
electroweak interaction.  Through the exchange of the massive $W$ and $Z$ vector bosons, the 
electrons and positrons comprising the positronium can produce a neutrino-antineutrino pair which
 can be accompanied by the emission of the photon off the incident electron or positron. To secure 
 the single photon total rate (here denoted by Ps $\rightarrow \gamma$), we compute the contributions 
 from each of the possible final states involving each of the three species of neutrinos 
 ($\nu_e, \nu_\mu, \nu_\tau$) and add the rates incoherently. For the final state containing the photon 
 plus the $\nu_e-\bar{\nu}_e$ pair, the relevant  Feynman diagrams with the final state photon emitted 
 off the incident electron leg are displayed in Fig. 1. In addition, there are the analogous graphs with the outgoing 
 photon emitted from the positron line which are not displayed. The process with the 
 $\nu_e \bar{\nu}_e \gamma$ final state includes both $t$-channel $W$ boson exchange (Fig. 1a) and 
 $s$-channel $Z$ boson exchange (Fig. 1b). In addition, the photon can also be emitted off the internal 
 charged $W$ line as shown in Fig. 2, but its contribution to the amplitude is further suppressed by 
 $\frac{m^2_e}{M^2_W}$, with $m_e$ and $M_W$ the electron and $W$ masses respectively, and is 
 thus neglected. For the $\nu_\mu \bar{\nu}_\mu \gamma$ and $\nu_\tau  \bar{\nu}_\tau \gamma$ 
 final states, only the $s$-channel $Z$ boson exchange displayed in Fig 1b contributes.  Here again, 
 there are the analogous graphs with the photon emitted off the positron line which are not explicitly 
 displayed. The $W$ exchange diagram of Fig. 1a does not contribute in either of these cases. The 
 energy-momentum of the incoming electron and positron are labeled as $p_{1}$, $p_{2}$ respectively, 
 while the outgoing neutrino and antineutrino are labeled as $p_{3}$ and $p_{4}$ respectively and  the
emitted photon 4-momentum is denoted by $k$. Here, the energy-momentum vectors are represented
 as $(E,\vec{p})$  and $\alpha \simeq $ 1/137 is the 
fine structure constant. We work throughout using natural units ($\hbar=c$ = 1).  Throughout the calculation, the tiny mass of the 
neutrinos is neglected and they are treated as massless. 
 
The decay rate of a bound positronium state into each particular mode, $\Gamma_{Ps\rightarrow\nu_l \bar{\nu}_l  \gamma }~~;~~ l=e, \nu, \tau$, can be expressed in terms of 
the cross section, $\sigma(e^+ e^- \rightarrow \nu_l \bar{\nu}_l \gamma)~;~ l = e, \mu, \tau$, associated
 with the process $e^= e^- \rightarrow \nu_l \bar{\nu}_l \gamma~~;~~l=e, \mu, \tau$ in which the
 $e^+, e^-$ constituents of the bound state as a free particles. So doing, the decay rate for the single photon 
 decay mode of Ps is then given by
 
\begin{eqnarray}
\label{eq-1}
\Gamma_{Ps\rightarrow \gamma }&=& \sum_{l=e, \mu, \tau}  \Gamma_{Ps\rightarrow\nu_l \bar{\nu}_l  \gamma } \cr
&=&\sum_{l=e, \mu, \tau}4v_{rel}|\Psi_{Ps}(0)|^{2}\sigma_{e^{+}e^{-}\rightarrow  \nu_l \bar{\nu}_l \gamma},
\end{eqnarray}

\noindent
where $|\Psi_{Ps}(0)|^{2}$ denotes the probability of finding 
the electron and positron in the same point of space and $v_{rel}$ 
is the relative velocity of the colliding $e^+ e^-$ pairs. Here the 
electron-positron annihilation cross sections 
$\sigma_{e^{+}e^{-}\rightarrow \gamma \nu_l \bar{\nu}_l}$ is 

\begin{eqnarray}
\label{eq-2}
\sigma_{e^{+}e^{-}\rightarrow \gamma \nu_l \bar{\nu}_l}=\frac{1}{4m_{e}^{2}v_{rel}} 
\int \frac{d^{3}p_{3}}{2(2 \pi)^{3}|\vec{p}_{3}|}\nonumber \\ 
\int \frac{d^{3}p_{4}}{2(2 \pi)^{3}|\vec{p}_{4}|}\int \frac{d^{3}k}{2(2 \pi)^{3}|\vec{k}|} \nonumber \\(2\pi)^{4}
 \delta^{(4)} \left(p_{1}+p_{2}-p_{3}-p_{4}-k\right)<|\mathcal{T}|^{2}>,  \nonumber \\
\end{eqnarray}

\noindent
where $<|\mathcal{T}|^{2}>=\frac{1}{4}\sum_{s_1, s_2, s_3, s_4}|\mathcal{T}|^{2}$ is the 
totally spin averaged square of the transition matrix element 
$\mathcal{T}$  obtained by averaging over the initial $e^-$ and $e^+$ 
spins ($s_1$ and $s_2$ respectively) and summing over the undetected
 final state neutrino and antineutrino spins ($s_3$ and $s_4$ respectively). 
 Eq. (\ref{eq-2}) includes the Lorentz invariant phase-space factors, as well as the 
energy-momentum conservation (which is guaranteed by the $\delta$ functions). 
Ps has a binding energy of 6.8 eV, which gives an upper bound to the 
momentum of the leptons involved in the decay of Ps. 
These values are negligible in comparison with the rest mass energy
 of those leptons, m$_{e}$. Thus, the initial 
 energy-momentum vectors are taken as $(m_{e},0,0,0)$. 
 Using this approximation and applying the momentum 
 conservation through the delta function, the integral over $d^3p_4$ in 
 Eq. (\ref{eq-2}) is performed, yielding
 
\begin{eqnarray}
\label{eq-3}
\sigma_{e^{+}e^{-}\rightarrow \gamma}=\frac{1}{4m_{e}^{2}v_{rel}2^{8}\pi^{5}} \int \frac{d^{3}p_{3}}{|\vec{p}_{3}|}
 \frac{1}{|\vec{p}_{3}+\vec{k}|}\int \frac{d^{3}k}{|\vec{k}|} \nonumber \\
 \delta \left(2m_{e}-|\vec{p}_{3}|-|\vec{p}_{3}+\vec{k}|-|\vec{k}|\right)<|\mathcal{T}|^{2}>.  \nonumber \\
\end{eqnarray}

\begin{figure}[h]
\includegraphics[width=8.cm]{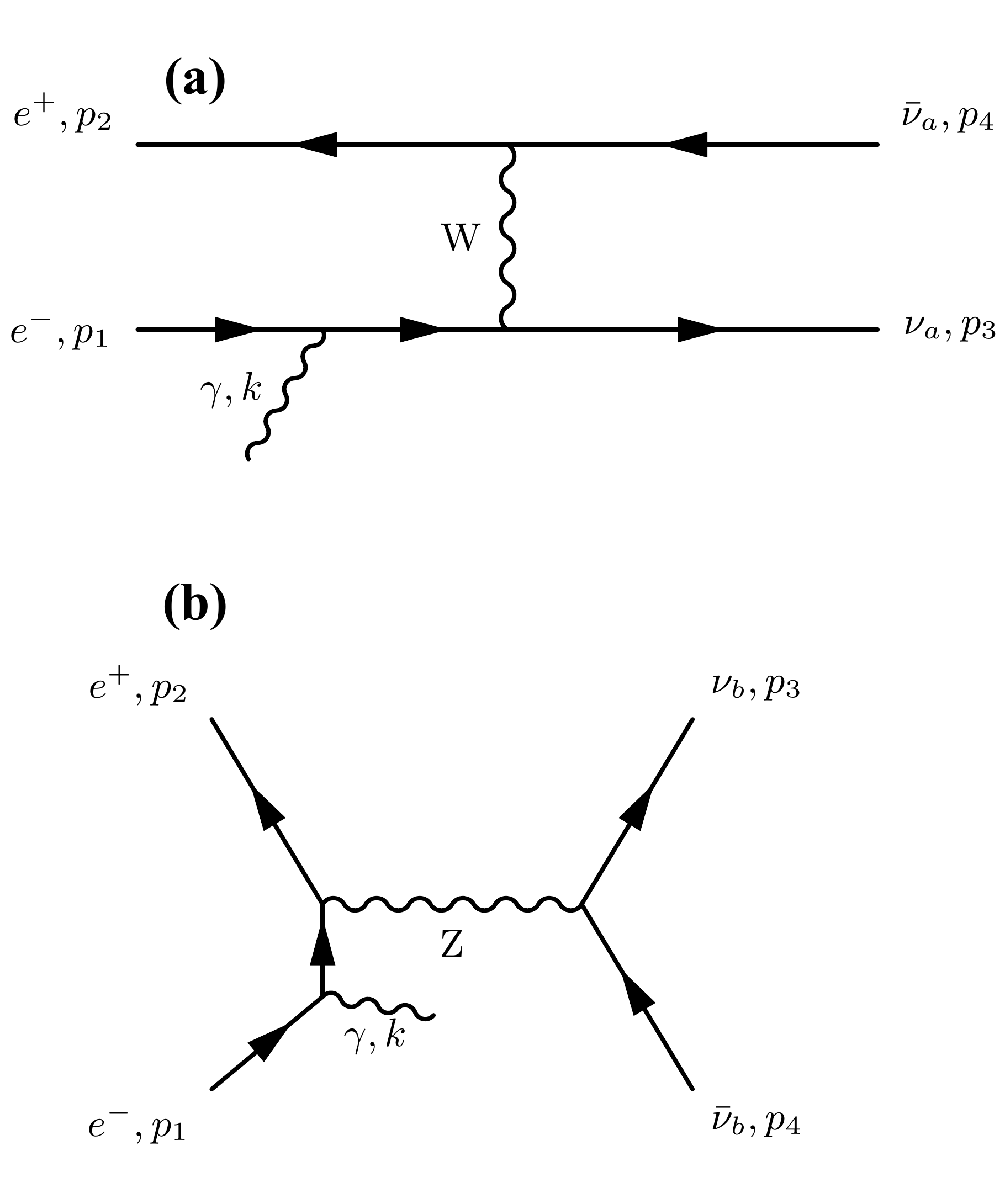}
\caption{Feynman diagrams contributions to the annihilation process 
e$^{+}$e$^{-}\rightarrow \nu_l \bar\nu_i\gamma$, with $l=e, \mu, \tau$. 
Both figures (a) and (b) contribute to the case $l=e$, while only figure 
(b) contributes to the cases $l=\mu, \tau$. In addition to these graphs, 
there are also the graphs where the outgoing photon is emitted off the 
positron leg. }
\label{default}
\end{figure}

\begin{figure}[h]
\includegraphics[width=8.cm]{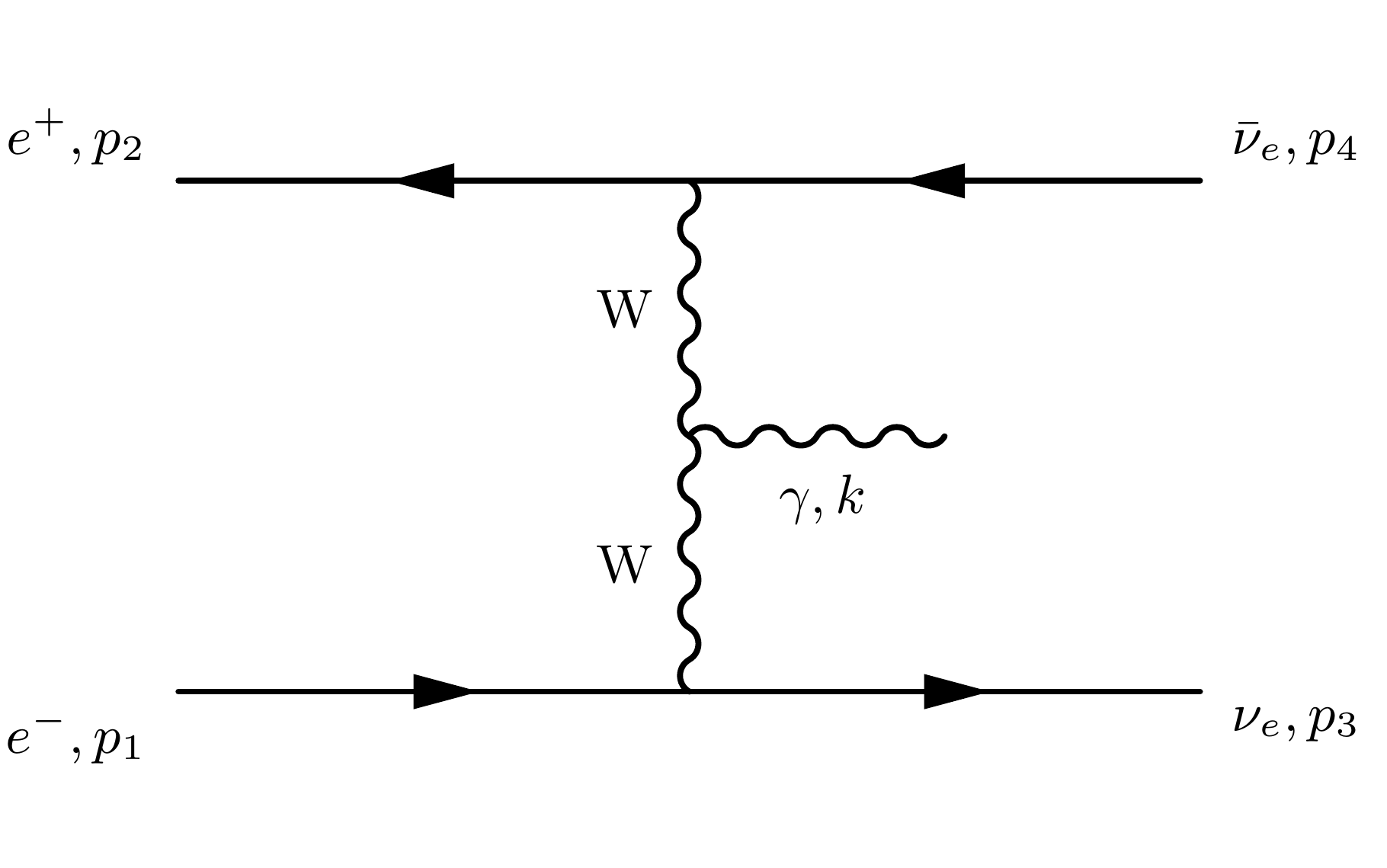}
 \caption{Feynman diagram contributing to 
 $\sigma(e^+ e^- \rightarrow \nu_e \bar\nu_e \gamma$ 
process which involves the exchange of two $W$ bosons. 
The contribution from this graph is thus highly suppressed
and is neglected in our analysis.}
\label{default}
\end{figure}

\noindent
This expression can be further simplified integrating over the 
angular degrees of freedom of $|\vec{p}_{3}|$ by virtue of the 
kinematic constrains present in the delta function resulting in

\begin{eqnarray}
\label{eq-4}
\sigma_{e^{+}e^{-}\rightarrow \gamma}=\frac{1}{4m_{e}^{2}v_{rel}2^{7}\pi^{4}} \int_{0}^{m_{e}} 
\frac{d^{3}k}{|\vec{k}|^{2}} \int_{m_{e}-|\vec{k}|}^{m_{e}}d|\vec{p}_{3}|
 <|\mathcal{T}|^{2}>.  \nonumber \\
\end{eqnarray} 

\noindent
Finally, the angular degrees of freedom of $\vec{k}$ are integrated, 
leading to the final form for the cross section for electron-positron
annihilation into one single photon

\begin{eqnarray}
\label{eq-5}
\sigma_{e^{+}e^{-}\rightarrow \gamma}=\frac{1}{4m_{e}^{2}v_{rel}2^{5}\pi^{3}} 
\int_{0}^{m_{e}} d|\vec{k}| \int_{m_{e}-|\vec{k}|}^{m_{e}}d|\vec{p}_{3}|
 <|\mathcal{T}|^{2}>.  \nonumber \\
\end{eqnarray} 

The relevant $T$ matrix element is secured from the Feynman 
graphs of Fig. 1 and the analogous graphs with the photon emitted off the positron leg. Since the momentum transfer is far smaller than 
the $W$ and $Z$ masses, it can be neglected in the $W$ and $Z$ 
boson propagators. So doing and working in Feynman gauge, one finds 
  
\begin{eqnarray}
\label{eq-7}
\mathcal{T}=\sum_{\lambda =1,2}\imath \frac{e^{3}G_{F}}{\sqrt{2}}\bigg[\delta_{l,e} \bar{v}(p_{2},s_{2})
\gamma^{\alpha} (1-\gamma_{5}) v_{e}(p_{4},s_{4}) \nonumber \\
\bar{u}_{e}(p_{3},s_{3})\gamma_{\alpha}(1-\gamma_{5})
\frac{\slashed{p}_{1}-\slashed{k}+m_{e}}{(p_{1}-k)^{2}-m_{e}^{2}}\gamma^{\nu} \epsilon^{*}_{\nu}(k, \lambda)
u(p_{1},s_{1}) \nonumber \\
+\bar{u}(p_{3},s_{3})\gamma^{\alpha}\frac{1-\gamma^{5}}{2}v(p_{4},s_{4}) \nonumber \\
\bigg(\bar{v}(p_{2},s_{2})(-\frac{1}{2}+\sin^{2}{\theta_{W}})\gamma_{\alpha}(1-\gamma_{5}) \nonumber \\
\frac{\slashed{p}_{1}-\slashed{k}+m_{e}}{(p_{1}-k)^{2}-m_{e}^{2}}\gamma^{\nu} \epsilon^{*}_{\nu}(k, \lambda)
u(p_{1},s_{1}) \nonumber\\
+\bar{v}(p_{2},s_{2})\sin^{2}{\theta_{W}}\gamma_{\alpha}(1+\gamma_{5}) \nonumber \\
\frac{\slashed{p}_{1}-\slashed{k}+m_{e}}{(p_{1}-k)^{2}-m_{e}^{2}}\gamma^{\nu} \epsilon^{*}_{\nu}(k, \lambda)
u(p_{1},s_{1})\bigg)\bigg]. \nonumber \\
\end{eqnarray} 

\noindent
 where the slash notation, {\it i.e.}, $\slashed{p}=\gamma^{\nu}p_{\nu}$ has been employed 
 and we have summed over the polarizations of the final state photon. 
Here $G_{F}= \frac{2\sqrt{2}\pi^2\alpha^2}{\sin^2\theta_W M_W^2}$  is the Fermi 
constant which has the numerical value 1.17$\times$10$^{-5}$ 
GeV$^{-2}$ and $\sin{\theta_{W}}\simeq 0.24$ is the weak mixing Weinberg  angle.
We employ the  Dirac $\gamma$ matrix conventions as is 
defined in Ref.~\cite{Peskin}. The diagrams with the outgoing photon emitted from the 
positron leg has the analogous structure except for the replacement the fermionic 
propagator $\frac{\slashed{p}_{1}-\slashed{k}+m_{e}}{(p_{1}-k)^{2}-m_{e}^{2}}$ with the propagator 
 $\frac{-\slashed{p}_{2}+\slashed{k}+m_{e}}{(-p_{2}+k)^{2}-m_{e}^{2}}$. The coefficient 
 $\delta_{l, e}$ guarantees that the $W$ exchange graph only contributes to the 
 $\nu_e \bar\nu_e \gamma$ final state.

With the amplitude $\mathcal{T}$ in hand, the spin averaged transition 
probability can be extracted by standard techniques. To do so, we employ 
 Mathematica~\cite{Mathematica}. However, before 
calculating the total cross section, let us first consider the differential cross 
section 

\begin{eqnarray}
\frac{d\sigma(e^+ e^- \rightarrow \gamma)} {d|\vec{k}|} &=& 
\sum_{l=e, \mu, \tau} \frac{d\sigma(e^+ e^- \rightarrow \nu_l \bar\nu_l \gamma)}
 {d|\vec{k}|} ,
\end{eqnarray}

\noindent
which in the present case represents 
the probability to have a collision per unit of area and energy of the 
emitted photon. It"s integral over the photon momentum magnitude gives the total cross section as 

\begin{eqnarray}
\label{eq-8}
\sigma_{e^{+}e^{-}\rightarrow \gamma}=\int_{0}^{m_{e}} d|\vec{k}| \frac{d\sigma (e^+ e^- \rightarrow \gamma)}{d|\vec{k}|}. 
\end{eqnarray} 

\noindent
The differential cross section is displayed in Fig. 3. The cross section is 
relatively flat as a function of the photon energy peaking when the energy
of the photon is 2m$_{e}$/3, which is represented as the red-dashed line
in Fig. 3. In other words, the most probable energy of the photon emitted 
will be 2m$_{e}$/3.

\begin{figure}[h]
\includegraphics[width=8.cm]{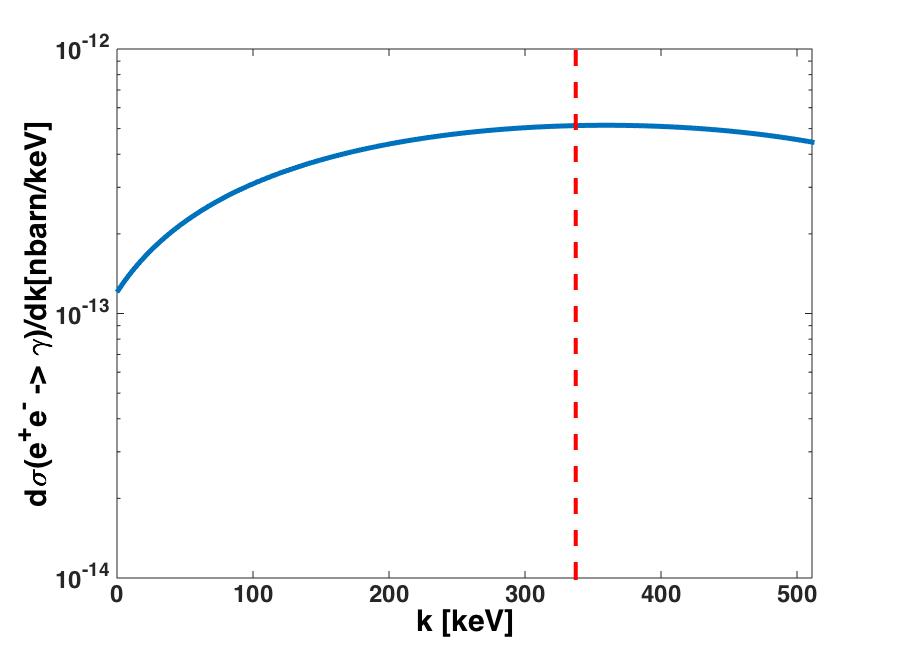}
\caption{Differential cross section for the process $e^{+}e^{-}\rightarrow \gamma$ as 
a function of the energy of the emitted photon. The red-dashed line represents the photon 
energy leading to the maximum probability of collision.}
\label{default}
\end{figure}

Finally, integrating the differential cross section from 
$[0,m_{e}]$ as instructed by Eq.(\ref{eq-8}) yields the total cross 
section of the decay mode e${+}$e$^{-}\rightarrow \gamma$ which 
is then employed in the calculation of the decay rate of Ps$\rightarrow \gamma$ yielding

\begin{eqnarray}
\label{eq-9}
\Gamma_{Ps\rightarrow \gamma }=|\Psi_{Ps}(0)|^{2}\frac{370.8\times G_{F}^{2}m_{e}^{2}e^{6}}{2^{7}\pi^{3}},
\end{eqnarray} 

\noindent
Using $|\Psi_{Ps}(0)|^{2}= m_{e}^3 \alpha^3 / (8\pi)$ gives a 
decay rate of $\Gamma_{Ps\rightarrow \gamma }$ = 1.72 $\times$ 10$^{-19}$ s$^{-1}$. 
This value is roughly the same order as for the $\nu \bar\nu$ (invisible) mode for o-Ps~\cite{Ps_decay}.

\section{Conclusions}

Due to the presence of the electroweak interactions, additional decay 
modes become available for the positronium atom. In the present work, 
we addressed what is effectively the single photon decay 
mode of parapositronium, which emerges from the 
$\nu_l \bar\nu_l \gamma~~;~~l=e, \nu, \tau$ final states and the non 
observation of the outgoing neutrinos and antineutrinos. These processes
 become operational due to the exchange of $W$ and $Z$ massive vector 
 bosons. The study of this decay aids in elucidating the role of the 
 electroweak interaction in the decay of Ps. Since this mode involves the emission of 
a single photon, it has a somewhat unique signature which in principle may be detected. The calculated 
decay rate is $\Gamma_{Ps\rightarrow \gamma}$ = 1.72 $\times$ 10$^{-19}$ s$^{-1}$,   
which is experimentally challenging, is comparable to the previously reported electroweak mediated decays in 
orthopositronium but without the photon emission. 

The decay mode presented will be also present in the decay of 
other exotic atoms, such as true muonium; the bound state of a muon 
and anti-muon. For this bound state,
the analogous single decay rate for  will be at 
enhanced by a factor of  $\sim 8 \times 10^{6}$  due to the heavier mass of the muon 
with respect to the electron. 

\section{Acknowledgements}

The work of J. P.-R. was supported by the Department of Energy, Office 
of Science, under Award Number DE-SC0010545.

%


\end{document}